\documentclass[twocolumn,pre,a4paper,showpacs,superscriptaddress]{revtex4-1}

\usepackage{amsbsy}
\usepackage{amssymb}
\usepackage[dvips]{graphicx}

\begin{document}

\title{Edge usage, motifs and regulatory logic for
cell cycling genetic networks}

\author{M. Zagorski}
\affiliation{Marian Smoluchowski Institute of Physics and Mark Kac Complex Systems Research Centre, Jagiellonian University,
Reymonta 4, 30-059 Krakow, Poland}%
\author{A. Krzywicki}%
\affiliation{Laboratoire de Physique Th\'eorique, Univ Paris-Sud ; CNRS, UMR8627, Orsay, F-91405, France}%
\author{O.C. Martin}%
\affiliation{INRA, Univ Paris-Sud, CNRS, UMR 0320 / UMR 8120 G\'en\'etique V\'eg\'etale, F-91190 Gif-sur-Yvette, France}%
\affiliation{Univ Paris-Sud, LPTMS ; CNRS, UMR 8626, F-91405, Orsay, France}

\date{\today}

\begin{abstract}
The cell cycle is a tightly controlled process, yet its underlying
genetic network shows marked differences across species. Which of the
associated structural features follow solely from the ability to
impose the appropriate gene expression patterns?
We tackle this question \emph{in silico} by examining the ensemble of all
regulatory networks which satisfy the constraint
of producing a given sequence of gene expressions. We focus on three cell
cycle profiles coming from bakers yeast, fission yeast and mammals.
First, we show that the networks in each of the ensembles use just a few
interactions that are repeatedly reused as building blocks. Second, we
find an enrichment in \emph{network motifs} that is similar in the two
yeast cell cycle systems investigated. These motifs do not have
autonomous functions, but nevertheless they reveal a regulatory
logic for cell cycling based on a feed-forward
cascade of activating interactions. 
\end{abstract}

\vspace{2pc}
\pacs{87.16.Yc, 87.18.Cf, 87.17.Aa}
\maketitle

\section{Introduction}
\par
The cell cycle -- biomass accumulation, DNA replication
and cell division -- is at the heart of life.
Disruptions in this cycling, due to environmental
changes or genetic defects, can lead to
cell death or to uncontrolled proliferation
such as in tumorigenesis~\cite{Malumbres_Barbacid_2009}.
It is thus not surprising that the cell cycle is
tightly controlled in all organisms, allowing a stereotyped
sequence of events to progress in an orderly and timely fashion.
Various molecular types participate in this orderly progress, but each
species has its own specific regulatory machinery
~\cite{Conlon_Raff_2003,Hochegger_Hunt_2008}.
\par
Detailed investigations along with reconstructions of network
interactions have allowed the building of
quantitative cell cycle models based on
ordinary differential equations (ODEs), in particular for
budding yeast~\cite{Chen_Calzone_Tyson_2004},
fission yeast~\cite{Sveiczer_Tyson_2004} and
mammals~\cite{Gerard_Goldbeter_2009}. The
mathematical complexity of these systems is very high.
To render them more comprehensible, simplifications 
have been proposed while preserving most of the qualitative
aspects of their dynamics~\cite{Csikasz-Nagy_Tyson_2006}.
\par
Reducing complexity still further, a number of authors 
have provided simplified models by discretising
time and by replacing the continuous concentrations of
molecular species by presence/absence (binary) variables.
Such a Boolean approach has a long history going back
to the pioneering work of Kauffman~\cite{Kauffman_1969} 
(see also the book~\cite{Kauffman_1993} and references 
therein) and since then it has been used in quite a number of systems.
In particular it has been applied to the cell cycle in
the three previously mentioned species~\cite{Li_Tang_2004, Lau_Tang_2007, 
Davidich_Bornholdt_2008, Faure_Naldi_Thieffry_2006, 
Boldhaus_Klemm_2010} with a fair amount of success. This Boolean approach 
is complementary to that based on ODEs: it is much less powerful 
quantitatively but it allows for simpler model construction and interpretation,
leading to qualitative insights into the generic aspects of regulatory rules.
The present work is of a similar vein, but uses a more realistic 
model of molecular interactions and focuses on the topology
of regulatory networks.
\par
Adjusting a dynamical model to reproduce observed cell cycle dynamics can 
be challenging but it is also an under-specified problem: because of the 
plethora of parameters, there are generally many different
solutions~\cite{Nochomovitz_Li_2006}. Certainly, nature takes advantage 
of this freedom, but a consequence is that related species can have quite 
different regulatory circuits, impeding attempts to extract common regulatory 
{\em principles}. To overcome this difficulty, we have taken an 
\emph{in silico} approach that characterizes all possible regulatory circuits 
that are compatible with given cell cycle expression dynamics. This route
allows us to determine which features are always or mostly present in these 
\emph{in silico} circuits and it also makes possible more meaningful comparative
studies of different biological networks.
\par
The present work builds on regulatory models developed in 
refs.~\cite{bkmz1,bkmz2} where we revealed several consequences 
of toy expression constraints on network structure. Here, 
we use the actual expression patterns found in three biological systems. 
Our modeling of the regulatory networks uses thermodynamic considerations 
to describe the underlying molecular interactions. Interestingly,
the actual number of effective inter-molecular 
interactions becomes an \emph{emergent} feature, driven
by biophysical constraints coupled to an 
appropriately defined fitness function mimicking the selection pressure 
operating in nature. Based on this choice of genetic regulatory network 
(GRN) modeling, we sample by Markov Chain Monte Carlo (MCMC) the space 
of all GRN that provide the proper gene expression patterns. We do 
so for the three ensembles of GRN where the gene expression profiles 
are taken from budding yeast, fission yeast and mammals derived in 
refs.~\cite{Li_Tang_2004, Davidich_Bornholdt_2008, Faure_Naldi_Thieffry_2006}. 
For each, we extract the structural properties of the sampled GRN, from 
edge usage to the presence of network motifs. Finally, 
by comparing to the case of an idealized cell cycle, we reveal an underlying 
common regulatory logic in which feed-forward activating cascades play a central 
role. This logic is apparent when the associated motifs are externally driven, but
the motifs on their own do not implement autonomous dynamics.
\par
In the next section we briefly sketch the main aspects of our model,
stressing the changes from our previous work~\cite{bkmz1,bkmz2}
to deal with new aspects of 
the problem. In section III we present our results.
We conclude in section IV.
\par
\section{Models and Methods}
\par
\subsection{Biophysical modeling of molecular interactions}
\par
In previous work~\cite{bkmz1,bkmz2} we developed a biophysical framework for
modeling the regulatory interactions between a transcription factor (TF)
and its DNA binding site. For cell cycling systems, a number of the 
actors, such as cyclins, are not TFs, so our previous framework must be 
generalized. The starting point is 
a network whose nodes and edges refer to molecules and activation/repression 
processes, be-they transcriptional, post-transcriptional, translational or 
post-translational. All interactions involve specific binding sites or surfaces
that allow two molecules to physically establish contact and bind
through the addition
of small forces. Almost all of the facing elements (atoms, bases, amino 
acids, ...) have to ``match'' for the two molecules to bind. Following 
ref.~\cite{Perelson_Weisbuch_1997}, we consider that each of the facing
regions is specified by an ordered 
lists of symbols that can be thought of as a string or a table 
characterizing that type of molecule and its site or surface
dedicated to that binding partner. Then the mutual binding energy of two 
molecules is taken to be additive in the number of mismatches between their 
strings, each mismatch contributing a penalty $\varepsilon$. 
\par
Such a framework is easily justified for TF-DNA binding energies and leads to 
a thermodynamic formula for the probability of occupation of a given DNA binding 
site in the presence of $n_j$ TFs of type $j$~\cite{Gerland_Hwa_2002}:
\begin{equation}
P_{ij} = \frac{1}{1 + 1/(n_j W_{ij})}
\, \, \, {\rm{where}} \, \, \,
W_{ij} = e^{-\varepsilon d_{ij}}.
\label{eq:probOccupy}
\end{equation}
Here $d_{ij}$ is the number of TF-DNA mismatches. The derivation of 
Eq.(\ref{eq:probOccupy}) in ref.~\cite{Hwa_Lectures_2000} shows that it is of 
rather general validity, so we shall apply it to all of the molecular species
in our cell cycle system.
\par
Note that $P_{ij}$ depends strongly on $d_{ij}$ and is appreciable only when 
$d_{ij}$ is small. With a realistic choice of model parameters, a small 
$d_{ij}$ is \emph{a priori} very unprobable, so that a functional 
molecular contact is also very unprobable.

\par
\subsection{Time dynamics of expression levels}
\par
Refs.~\cite{bkmz1,bkmz2} considered \emph{transcriptional} regulation; we 
have to generalize the framework to any kind of regulation because of the 
different molecular types driving the cell cycle. Let $S_j(t) \in [0,1]$ 
($j=1,\ldots N$) denote the average level of expression of the molecular 
species $j$ at time $t$, normalized by its maximum value. 
The $N$ expression levels $\{S_j(t)\}$ will be referred to as the 
{\em  phenotype} at time $t$. We are interested in processes of the type
\begin{equation}
\{S_j(t_0)\} \to \{S_j(t_1)\} \to \ldots \to \{S_j(t_M)\}
\end{equation}
that must reproduce as well as possible a target cell cycle sequence, 
\emph{i.e.}, the sequence deduced from experimental measurements. For example, 
in the case
of the fission yeast \emph{S. Pombe}, this sequence pattern is given in 
Table \ref{tab:pmb} as specified in ref.~\cite{Davidich_Bornholdt_2008}. 
\begin{table}[h]
\centering
\begin{tabular}{cccccccccc}
 \text{Start} & \text{SK} & \text{Cdc2} &
\text{Cdc25} & \text{Cdc2*} & \text{Slp1} &
   \text{PP} & \text{Ste9} & \text{Rum1} & \text{Wee1} \\
 1 & 0 & 0 & 0 & 0 & 0 & 0 & 1 & 1 & 1 \\
 0 & 1 & 0 & 0 & 0 & 0 & 0 & 1 & 1 & 1 \\
 0 & 0 & 0 & 0 & 0 & 0 & 0 & 0 & 0 & 1 \\
 0 & 0 & 1 & 0 & 0 & 0 & 0 & 0 & 0 & 1 \\
 0 & 0 & 1 & 1 & 0 & 0 & 0 & 0 & 0 & 0 \\
 0 & 0 & 1 & 1 & 1 & 0 & 0 & 0 & 0 & 0 \\
 0 & 0 & 1 & 1 & 1 & 1 & 0 & 0 & 0 & 0 \\
 0 & 0 & 0 & 1 & 0 & 1 & 1 & 0 & 0 & 0 \\
 0 & 0 & 0 & 0 & 0 & 0 & 1 & 1 & 1 & 1 \\
 0 & 0 & 0 & 0 & 0 & 0 & 0 & 1 & 1 & 1
\end{tabular}
\caption{The cell cycle expression profiles of fission yeast (from~\cite{Davidich_Bornholdt_2008}). 
Time runs from the top to the bottom. Successive, idealized expression 
levels, are either 0 or 1, at each time step. We keep the terminology from ~\cite{Davidich_Bornholdt_2008}: the initial state's expression is given by 
the first line of the table (referred to as a ``START'' state) and the last state is a fixed point associated with the cell size check point. When a critical cell size is reached, a signal (implemented by having
a so-called ``Start'' node turn on) triggers the cycle again. The Cdc2 (Cdc2*) stands for Cdc2/Cdc13 (Cdc2/Cdc13*), where the * sign indicates the highly activated form of
the complex~\cite{Davidich_Bornholdt_2008}. Similarly Wee1 stands for the Wee1/Mik1 kinases.}
\label{tab:pmb}
\end{table}
\par
Following refs.~\cite{bkmz1,bkmz2}, the discrete time dynamics
for the expression levels are based on a mean field approximation
which neglects cooperative binding effects. Explicitly, we have
\begin{equation}
S_i(t+1) = \{1 - \prod_{j} [1 - P_{ij}(t)] \} \prod_{j'} [1 - P_{ij'}(t)]
\label{eq:inhibitors}
\end{equation}
where the $P_{ij}$ were defined in Eq.(\ref{eq:probOccupy}). In the present
expression, $j$ runs over activating and $j'$ over
inhibitory interactions respectively. It is easy to see
that Eq.(\ref{eq:inhibitors}) embodies a 
simple logic for each $S_i$: (i) each activator or repressor
acts independently; (ii) the binding of at least
one activator is required for activation of the target;
(iii) the binding of even a single repressor is sufficient to veto the
activation and so will turn off the expression of the target. For 
the sake of simplicity we set
$n_j= nS_j(t)$ in Eq.(\ref{eq:probOccupy}),
with a value of $n$ common to all molecular
species. For our simulations, we set $n=1000$, but we have checked that the
results presented depend little on the
magnitude of this parameter. Thus the fluctuations in the number
of molecules of a given type will not matter much, further justifying
our mean field approach.
\par
Notice that Eq.(\ref{eq:probOccupy}) holds provided
the system is in equilibrium. Thus, in writing
Eq.(\ref{eq:inhibitors}) it is assumed that
the binding-unbinding reaction is fast compared
to that for full activation/repression. However,
the rate of a binding-unbinding reaction is proportional
to the concentration of reactants and the
equilibration time becomes very large for small concentrations.
To avoid this limitation, attempts to generate very small 
concentrations should be ignored. Furthermore, the unphysical assumption that there is always enough time to reach equilibrium sometimes leads to pathologies. In particular, the system suffers from an instability: even a relatively weak interaction has an \emph{a priori} capacity of generating a spurious progressive amplification of physically unsignificant (because very low) expression levels. To avoid this bad behavior and improve the robustness of the model, we have recourse to a phenomenological correction,
introducing a threshold $H$ to modify Eq.(\ref{eq:inhibitors})
whenever the expression level is too small:
\begin{equation}
S_i(t+1) = S^{min} \; \; \mbox{\rm when}   \;  \; S_i(t+1) <  H
\label{eq:dynamics}
\end{equation}
In effect, $S^{min}$ can be thought of imposing a basal rate or
some level of leakiness on the transcription while the threshold $H$ ensures that
expression levels stay low unless the input signal is sufficiently
activatory. In practice
we set $S^{min} = 1/n = 0.001$ as if there
were just one molecule of that species and $H$ is set to
$H=0.01$. Then, $S_i(t)$ can leave the minimal level
only when at least one $d_{ij}$ is small enough to have
a dynamical significance~\cite{bkmz2}.
\par
\subsection{Computational implementation}
For each $i$ and each $j$, $(i,j)$, the interaction strength $W_{ij}$ is given by Eq.(\ref{eq:probOccupy}) where $d_{ij}$
can be considered to be the mismatch~\cite{Perelson_Weisbuch_1997} between two strings of length $L$.
In the simulation, rather than storing the 2 character strings for each such
interaction, we simply store the binary difference string (also of length $L$) specifying which entries
match (1) and which mismatch (0). There is also a sign associated with each interaction: + for an activator, - for a repressor. Mutations can change the sign of an interaction or they can transform a match into a mismatch and vice versa. Following the procedure used for TFs~\cite{bkmz1,bkmz2}, and to keep the modeling simple, we consider that the original strings use a four letter code. Then the probability that a mutation replaces a match by a mismatch is 3/4, while the probability that a mismatch is converted into a match is 1/4, embodying the fact that it is easier to have a mismatch than a match. (Note that even for protein-protein interactions, this rule can be motivated by the fact that molecular changes are often the result of a point mutation at the DNA level.) We will refer to the matrix $d_{ij}$ and to the associated set of signs as the \emph{genotype}. Typically $d_{ij} \neq d_{ji}$ since these two numbers of mismatches correspond to different processes: the former to the activation of $i$ by $j$, while the latter to the activation of $j$ by $i$. In these two processes different active sites have to match.
\par
For a given genotype, the dynamical system modeling
the cell cycle is initialized in the ``START'' 
state and then the trajectory is generated by iterating Eq.(\ref{eq:dynamics}). 
At each time step we compute the distance $D(t)$
\begin{eqnarray}
D(t) = \sum_{i=1}^N |S_i(t) - S_i^{target}(t)|
\end{eqnarray}
between the vectors of expression levels associated
with the actual sequence of the dynamical system
at hand and the target one (for fission yeast, the
target sequence is given in Table~\ref{tab:pmb}).
The total distance $D_T=\sum_{t=0}^{T} D(t)$ is
then used to define the {\em fitness} $F$ of the genotype via
\begin{eqnarray}
F = e^{-f D_T}
\label{eq:fitness}
\end{eqnarray}
where the parameter $f$ gives the strength of the
selection pressure to maintain ``good'' expression profiles. 
Indeed, deviations from the target expression profile are
likely to be deleterious for the proper functioning of a
cell and its progeny. Note that this
fitness defines a weight for each genotype and thus
an ensemble where fitness plays the role of a Boltzmann factor.
Finally, to sample genotypes according to their
fitness, we use Markov Chain Monte Carlo (MCMC).
The fitness enters the associated Metropolis test which accepts or
rejects a trial change (mutation) to the genotype.
We always start with a completely random network and
thermalize it until it has a good fitness. Then,
we produce a long sequence of several thousand GRN genotypes, 
each being separated by 200 sweeps (a sweep is a
series of $N^2 L$ attempted mutations and sign flips). To deal with possible fragmentation of the search space~\cite{Boldhaus_Klemm_2010}, we checked that the statistical properties of the resulting genotypes were the same for 
several independent simulations, \emph{i.e.} each simulation
being initialized with a different random genotype.
\par
The parameters values used in our computations are $L=12$,
$\varepsilon=1.75$, $f=20$, $H=0.01$ and $n=1000$.
Compared to ref.~\cite{bkmz2},
$H$ is a new parameter, $\varepsilon$ is slightly smaller while
the remaining are the same. Interestingly, it is essential that $L$
and $\varepsilon$ be such that the \emph{a priori} probability
of a small mismatch is very low. Once that constraint is
taken into account, the model
results are robust with respect to the variation
of the parameters. 
\par
The Metropolis test described above forces 
$D_T$ to be relatively small if $f$ is large, but the MCMC is inefficient
in that regime, forcing us to work with not too large values of $f$. Then  
in the $T\times N$ table of numbers $S_i(t)$, there may appear 
a few that are ``ambiguous'', that is that are away from the target 
value which in this study is either 0 or 1. For
example an expression level may have the value 0.3 in a place where the
target value is 0. Forcing a better agreement by 
increasing $f$ is not computationally feasible. 
To overcome this difficulty, from our large MCMC sample, we have selected
genotypes where such ambiguities are absent, imposing that expressed 
(respectively unexpressed) molecular types have in fact
levels above $0.6$ (respectively below $0.2$). This 
selection is done mostly 
for esthetic reasons, the properties of the full sample
are essentially the same because of the low frequency of these ambiguities.
\par
\section{Results}
\subsection{Essential interactions} 
\par
The networks generated by the MCMC are characterized
at a microscopic level by weights
$W_{ij}$ quantifying the regulatory interaction strength of
a molecule of type $j$ on the expression level of
molecules of type $i$. Just as had been found in our
previous work~\cite{bkmz1,bkmz2}, most $W_{ij}$ remain 
negligible, but a small
fraction have values that are far above the
background level. These interactions are
the ones important for reproducing the target sequence of expression
states. We formalize this property
as follows: if setting $W_{ij}=0$ causes the
Metropolis test to reject the change
5 times in a row, we say that the interaction $W_{ij}$ is ``essential''.
The set of essential interactions of a network allows for
a \emph{summary} representation that
is well suited for visualising the influences of each molecular type.
A genotype's set of essential interactions defines a
directed network, the {\em essential network}, from which one
may extract informative regulatory patterns.
\par
\begin{figure}
\includegraphics[width=8.2cm]{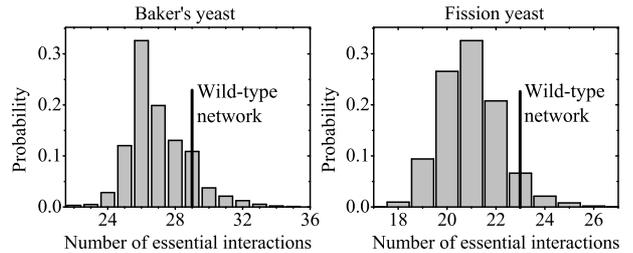}
\caption{Number of essential interactions in regulatory networks 
generated within the baker's (left) and fission (right) yeast 
ensembles. In both 
cases the number of interactions measured in the 
wild-type networks~\cite{Li_Tang_2004, Davidich_Bornholdt_2008} 
is close to the position of the peak in the ensemble's distribution.}
\label{fig:YST_PMB_ess}
\end{figure}

\begin{table}
\begin{tabular}{lcccc} \hline \hline
Observable & Baker's& Fission& Mammalian& Idealized\\ \hline
$E_{wt}$                   & 29        & 23       & 22        & --- \\
$\langle E_{ess} \rangle$   & 27.98(2)  & 20.96(2) & 13.92(1)  & 17.62(2) \\
$\langle E_{rep} \rangle$   & 12.20(2)  & 9.02(2)  & 5.30(2)   & 6.78(1) \\
$\langle D_T     \rangle / T$   & 0.534(1)  & 0.493(1) & 0.491(1)  & 0.358(1) \\
$\langle S_{ON}  \rangle$   & 0.881(1)  & 0.861(1) & 0.879(1)  & 0.920(1)\\
$\langle S_{OFF} \rangle$   & 0.011(1)  & 0.014(1) & 0.026(1) & 0.006(1)  \\
\hline \hline
\end{tabular}
\caption{Statistics from networks associated with the
cell cycle expression patterns of baker's
yeast, fission yeast, mammals, and with an idealized cell cycle
expression pattern. $E_{wt}$ stands for number of
interactions (edges) in the wild-type network. The other 
symbols refer to properties of \emph{in silico}
networks generated by MCMC: $E_{ess}$ is the number 
of essential interactions; $E_{rep}$ is the number 
of inhibitory interactions; $D_T$ is the total 
distance between the actual expression patterns and target one
(nearly constant 
and $\approx 0.5$ when divided by the number of steps of the cycle);
$S_{ON}$ ($S_{OFF}$) is the expression level of genes that are ON
(OFF) in the target phenotype.}
\label{tab:stats}
\end{table}
It is well known that biological GRNs are sparse. As shown in
refs.~\cite{bkmz1,bkmz2}, our model very naturally generates
sparse essential networks because 
relevant -- \emph{i.e.} small -- mismatches are 
\emph{a priori} unprobable and arise only
under strong selection pressure. Then simply because of
entropy, the model strongly favors low
numbers of essential interactions. However, 
this argument is qualitative only. Let us thus 
compare quantitatively the numbers of edges produced within the model to
the numbers in the biological data as inferred by the
different authors~\cite{Li_Tang_2004,Davidich_Bornholdt_2008,Faure_Naldi_Thieffry_2006}.
Hereafter, we will refer to those reconstructed networks
as the ``wild-type'' networks. In Table~\ref{tab:stats} 
we give the average number of
essential interactions along with the value in the wild-type 
network for baker's yeast~\cite{Li_Tang_2004}, fission 
yeast~\cite{Davidich_Bornholdt_2008} 
and mammals~\cite{Faure_Naldi_Thieffry_2006}. (Note: we
do not count the ``self-degradation'' interactions added by hand 
by some of these authors which are specific
to their modeling of the discrete time dynamics of
the expression levels.) The mammalian wild-type network appears 
to use significantly more interactions than our model but the two yeast species seem to be in good agreement with our model.
To quantify this, we have measured the fluctuations in each ensemble.
As shown in Fig.~\ref{fig:YST_PMB_ess}, for the two yeast ensembles 
the distributions of the number of edges
are bell shaped and narrow, so in fact the
fluctuations are very mild. Furthermore, the difference between
the wild-type numbers and the expected value in these ensembles
are not much more than one standard deviation, a result that
is quite non-trivial.
Most of results presented hereafter concern the two
species of yeast. The fact
that our model does much better for yeasts than for mammals may not be
a coincidence. Indeed, although baker's yeast and fission yeast
are highly divergent evolutionarily, much more so than
any mammals amongst themselves, they are both uni-cellular
organisms. It is well known that the regulatory control of 
gene expression is intrinsically more complex in multi-cellular
organisms, so \emph{a posteriori} one may draw some satisfaction
from the fact that our simple model seems to do well when
regulation is simple but does much less well when regulation 
involves very complex mechanisms. In the remaining subsections we will omit results referring to the mammalian cell cycle. It is sufficient to say that they are systematically off the data.

\subsection{Edge usage} 
\par
\begin{figure}
\includegraphics[width=8.6cm]{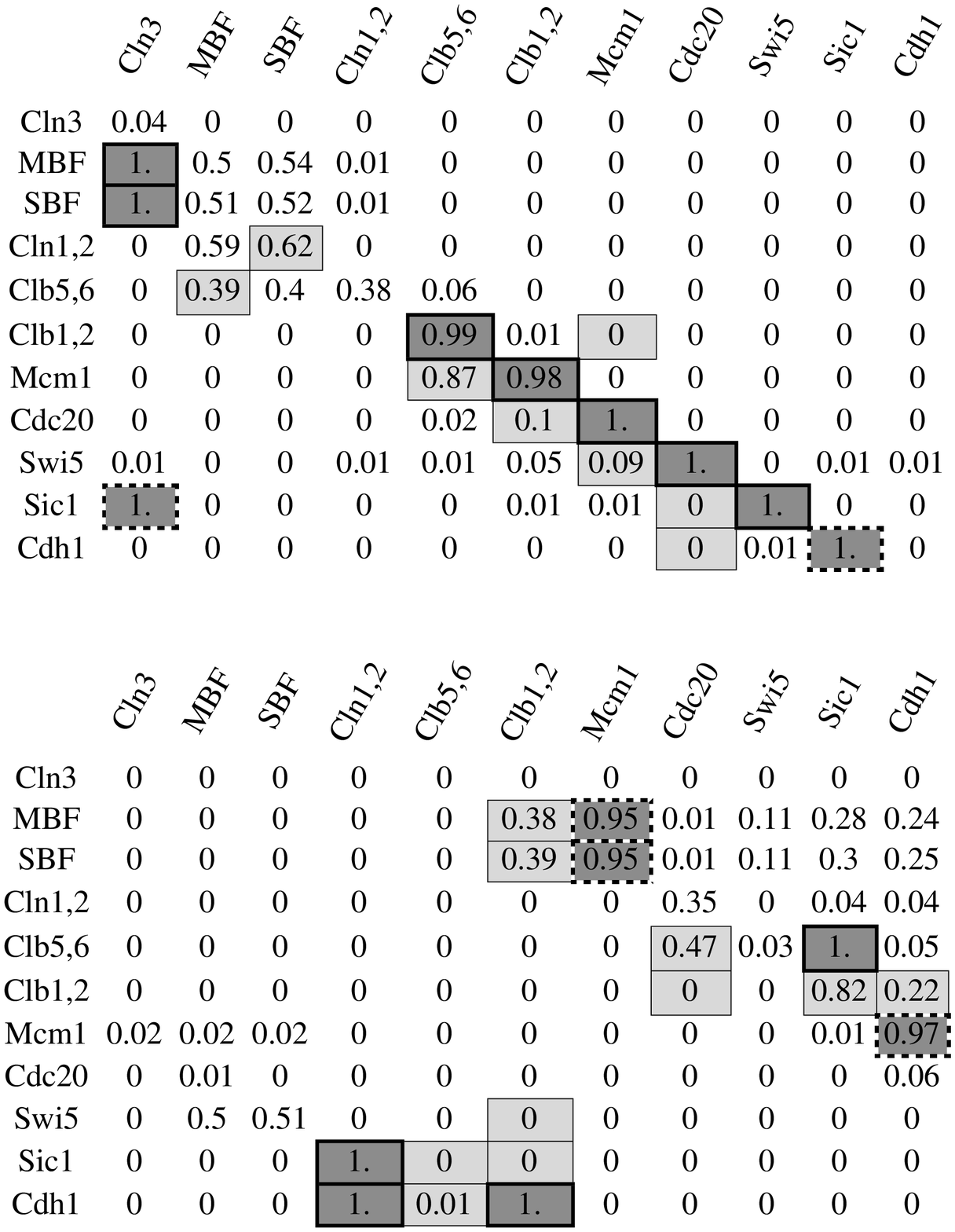}
\caption{Frequencies of activators (top) and inhibitors (bottom) for
networks produced in the ensemble relevant for 
the cell cycle in baker's yeast. Array element $(i,j)$ corresponds to the
probability of finding a link $i \leftarrow j$ in the analysed ensemble of GRNs. 
Dark grey highlighted entries have frequencies 
higher than 90$\%$; a thick solid (respectively thick dashed) 
frame indicates that the link is present (respectively absent) in the
yeast wild-type network 
of ref.~\cite{Li_Tang_2004} (see also Fig.~\ref{fig:YST_GRNs}). Light 
grey fields indicate the other activators/repressors
in the wild-type network.}
\label{fig:freq_YST}
\end{figure}

Now consider the edge usage in the essential networks. The frequency with which
there is an essential interaction from node $j$ to node $i$ can be represented
by a square matrix with entries between 0 and 1. We have computed these matrices
for the three cell cycles, and in Fig.~\ref{fig:freq_YST} we display the results
separately for activating and inhibitory interactions in baker's yeast. Particularly, there are 9 activators (7 common with the wild-type) and 7 repressors (4 common with the wild-type) almost always present in the ensemble for baker's yeast (\emph{cf.} dark grey entries in Fig.~\ref{fig:freq_YST}). Further, the results can be compared with frequencies of activatory/inhibitory interactions from~\cite{Lau_Tang_2007}, where 6 activators and 4 repressors (all common with the wild-type) are found to be absolutely required for a network to produce the cell-cycle process. Specifically, among these absolutely required interactions, 8 edges are almost always present and two edges are present in more than 80\% of networks in our ensemble.
\par
On a more general level there is a manifest dichotomy: certain interactions 
are very rarely if ever present,
while others are often if not always present. Furthermore, there is a clear structure
to the matrix: in the line
just below the main diagonal, activating
interactions are very frequent.
A different pattern arises for the inhibitory interactions: there, all the
entries near the diagonal are always absent, and elements 4 or 5 steps
shifted from the diagonal are frequently or 
always present. 

\begin{figure}
\includegraphics[width=8.6cm]{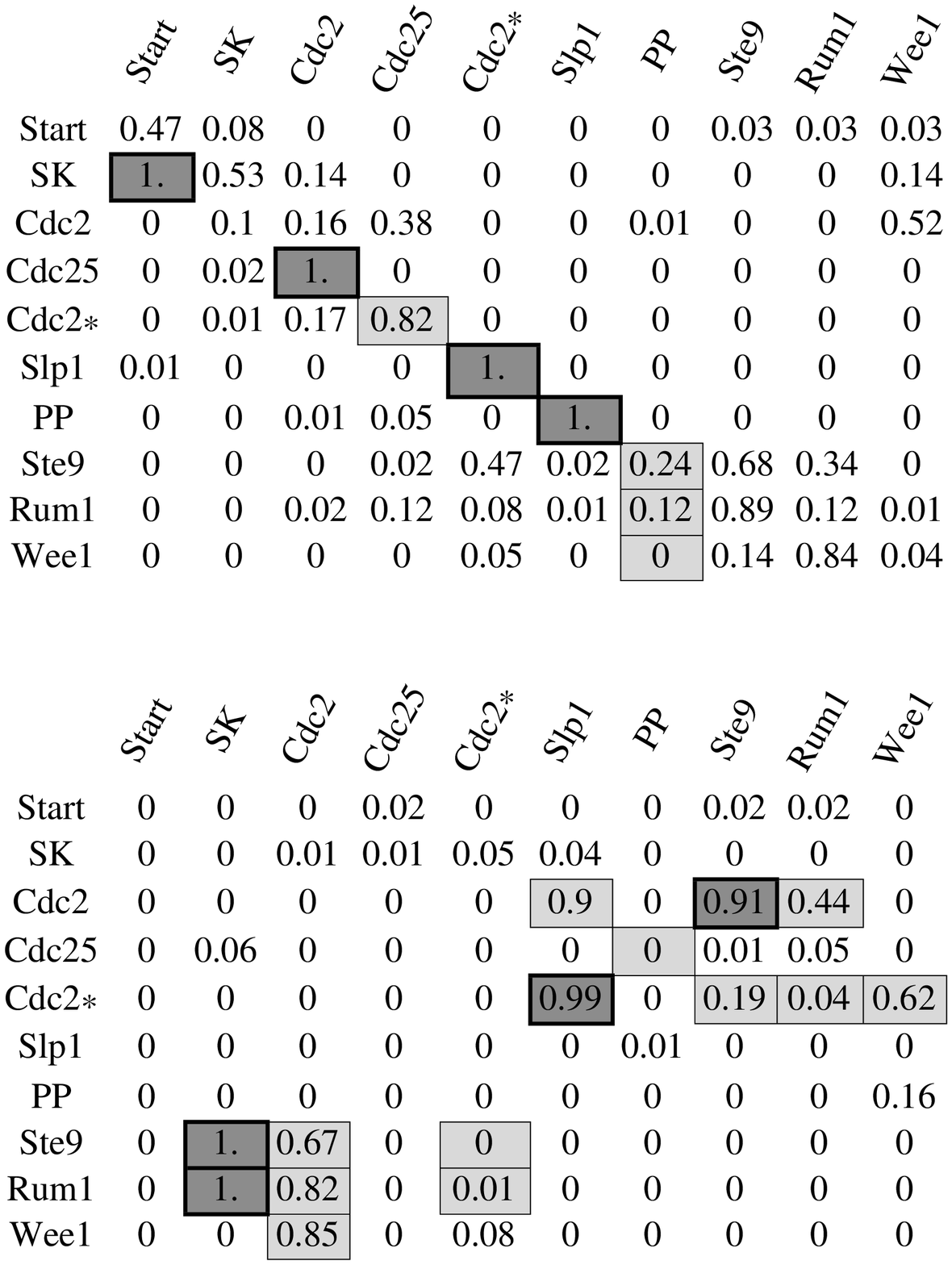}
\caption{Frequencies of activators (top) and inhibitors (bottom) for 
networks produced in the ensemble relevant for
the cell cycle in fission yeast. The conventions
for highlighting the values are the same as in 
Fig.~\ref{fig:freq_YST} but here refer to the wild-type network 
of fission yeast as specified in ref.~\cite{Davidich_Bornholdt_2008}
(see also Fig.~\ref{fig:PMB_GRNs}).}
\label{fig:freq_PMB}
\end{figure}

Qualitatively the same pattern arises
in fission yeast but it is noisier (Fig.~\ref{fig:freq_PMB}), and in
the case of the mammalian cell cycle which involves only 7 genes,
there are simply remanants of these patterns.

\subsection{Overlaps with the wild-type networks} 
\par
\begin{figure*}
\includegraphics[width=17.6cm]{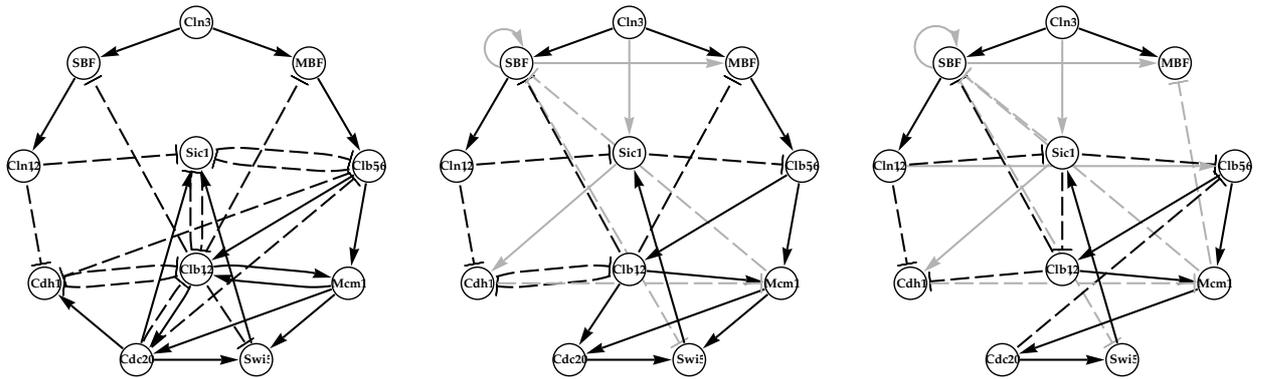}
\vspace{-0.8cm}
\caption{Left: The wild-type regulatory network for the baker's yeast cell cycle has 
29 links, of which 15 are activating (solid) and 14 are inhibitory (dashed). 
Compared to ref.~\cite{Li_Tang_2004} we do not include the 
added self-degradation links. 
Middle: Regulatory network in the MCMC ensemble with the highest fraction 
of hits (73\% of the essential interactions are present in the wild-type).
This network has 19 hits (black) and 7 non-hits (light grey) for a total of 
26 essential interactions. 
Right: The most frequent network in the baker's yeast ensemble (2.6\% of all 
GRNs, 16 hits (black) and 10 non-hits (light grey) for a total of 26 essential
interactions).}
\label{fig:YST_GRNs}
\end{figure*}
\begin{figure*}
\includegraphics[width=17.6cm]{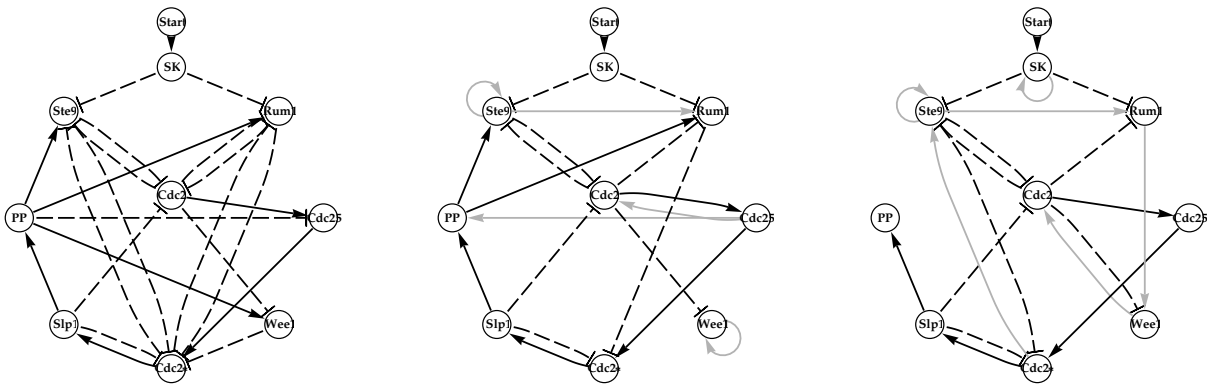}
\vspace{-0.8cm}
\caption{Left: The wild-type regulatory network for the fission yeast cell cycle has 23 
links, of which 8 are activating (solid) and 15 inhibitory (dashed). 
Compared to~\cite{Davidich_Bornholdt_2008} we do not include the added 
self-degradation links. 
Middle: Regulatory network in the MCMC ensemble 
with the highest fraction of hits (76\% of the essential interactions 
are present in the wild-type).
The network has 16 hits (black) and 5 
non-hits (light grey) for a total of 21 essential interactions. 
Right: Most 
frequent network in the fission yeast ensemble (2.1\% of all GRNs, 14 hits (black)
and 6 non-hits (light grey) for a total of 20 essential 
interactions).}
\label{fig:PMB_GRNs}
\end{figure*}

If an essential interaction of an \emph{in silico} GRN is  
present in the wild-type network,
we shall shall call this interaction a ``hit''.
What fraction of the essential interactions of a GRN are hits? 
If the fraction is close to 1, there is a very strong
overlap between that GRN and the wild-type network.
\begin{figure}
\includegraphics[width=8.2cm]{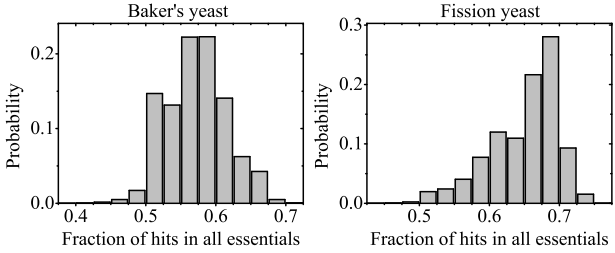}
\caption{Histograms for the
fraction of essential interactions that are ``hits'' in the baker's (left) 
and fission (right) yeast ensemble. We see that networks in our ensembles have 
roughly 50\% to 70\% of interactions in common with the wild-type networks. Notice, that the wild-type networks are not part of our ensemble.}
\label{fig:YST_PMB_frac}
\end{figure}
As can be seen in Fig.~\ref{fig:YST_PMB_frac}, in practice
the fraction varies from network to network but 
typically takes values in the range from 50\% to 70\% in
both yeast species. The \emph{mean} fractions are 65\% and 57\% for 
fission and baker's yeast so clearly there is a strong overlap
between the \emph{in silico} and wild-type networks. The highest
fraction of hits found in our MCMC is 73\% for baker's yeast
and 76\% for fission yeast.  Notice that this fraction is close to the overlap of hits with respect to all links in the wild-type network, since in our  model $E_{wt}$ and $E_{ess}$ are similar for both yeast species (see Tab.~\ref{tab:stats}). To compare visually the associated networks
to the corresponding wild-type networks, we display them in
Fig.~\ref{fig:YST_GRNs} for baker's yeast and in Fig.~\ref{fig:PMB_GRNs} 
for fission yeast.
The main differences arise for the Sic1 and Clb1,2 genes
which in the wild-type have higher degree and more mutually
inhibitory interactions than in the GRN generated by MCMC.
Such interaction pairs are quite natural in the context
of protein-protein interactions and thus probably reveal
a shortcoming of our modeling framework. 

\subsection{A multitude of different essential networks} 
\par
\begin{figure}
\includegraphics[width=6.0cm]{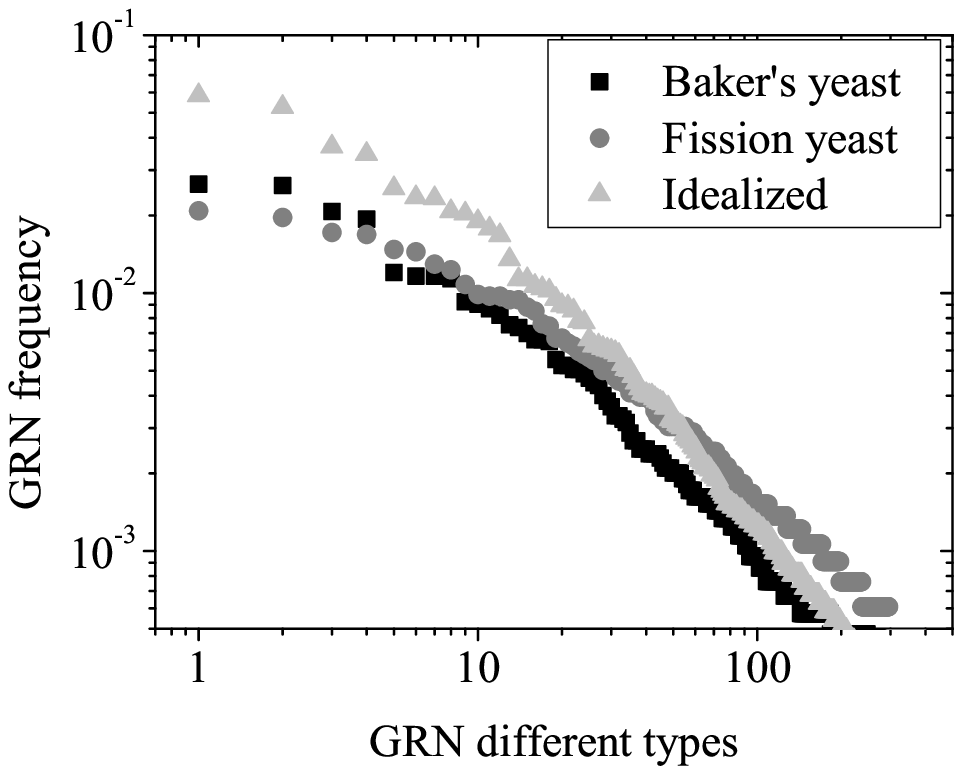}
\caption{Frequency in the MCMC of the different
essential networks for three ensembles: 
baker's yeast (squares), fission yeast (circles) and idealized cell-cycle 
(triangles). In all cases there are a few most frequent networks 
and the distribution of frequencies for the rarer 
networks roughly follows a power law.}
\label{fig:net_frequency}
\end{figure}

The MCMC generates a very large number of genotypes from
which we construct essential networks. In effect, each
essential network corresponds
to a different way interactions can be specified so as to 
reproduce the target gene expression dynamics.
Many genotypes give rise to the same essential network,
but are there many different essential networks? The answer
of course is yes, we find hundreds of different
such networks. One may then ask whether some 
arise much more frequently than others. On the right
hand side of 
Fig.~\ref{fig:YST_GRNs} (respectively Fig.~\ref{fig:PMB_GRNs})
we show the most frequently found essential network
for baker's yeast (respectively fission yeast); these
arise with frequencies between 2 and 3\%. Their
characteristics are similar to those of the essential networks
having the highest overlaps with the wild-type networks. 
These most frequent essential networks arise hundreds
of times more frequently than the rarest
ones. In Fig.~\ref{fig:net_frequency} we show the associated
rank histogram for the concurrence frequencies on a log-log plot. The fat
tail in this distribution is roughly compatible with a power law. Such 
a shape for a rank histogram arises in a number of other systems,
in particular in neutral networks where many genotypes
give rise to the same phenotype.
%

\par
\subsection{Network motifs}
In the terminology of Alon~\cite{Alon_book_2007}, 
a network motif in a (biological) graph is a subgraph that
is present at a higher frequency than 
expected at random. The term ``at random''
is generally defined using an ensemble of graphs
where each node is constrained to have the same in and 
out degree as in the biological graph of interest. In practice
this ensemble is generated  using a randomization of the
edges of the biological graph~\cite{MiloShen-Orr2002, MaslovSneppen2002} and that is
what we have done too.

\begin{figure}[ht]
\includegraphics[width=7.0cm]{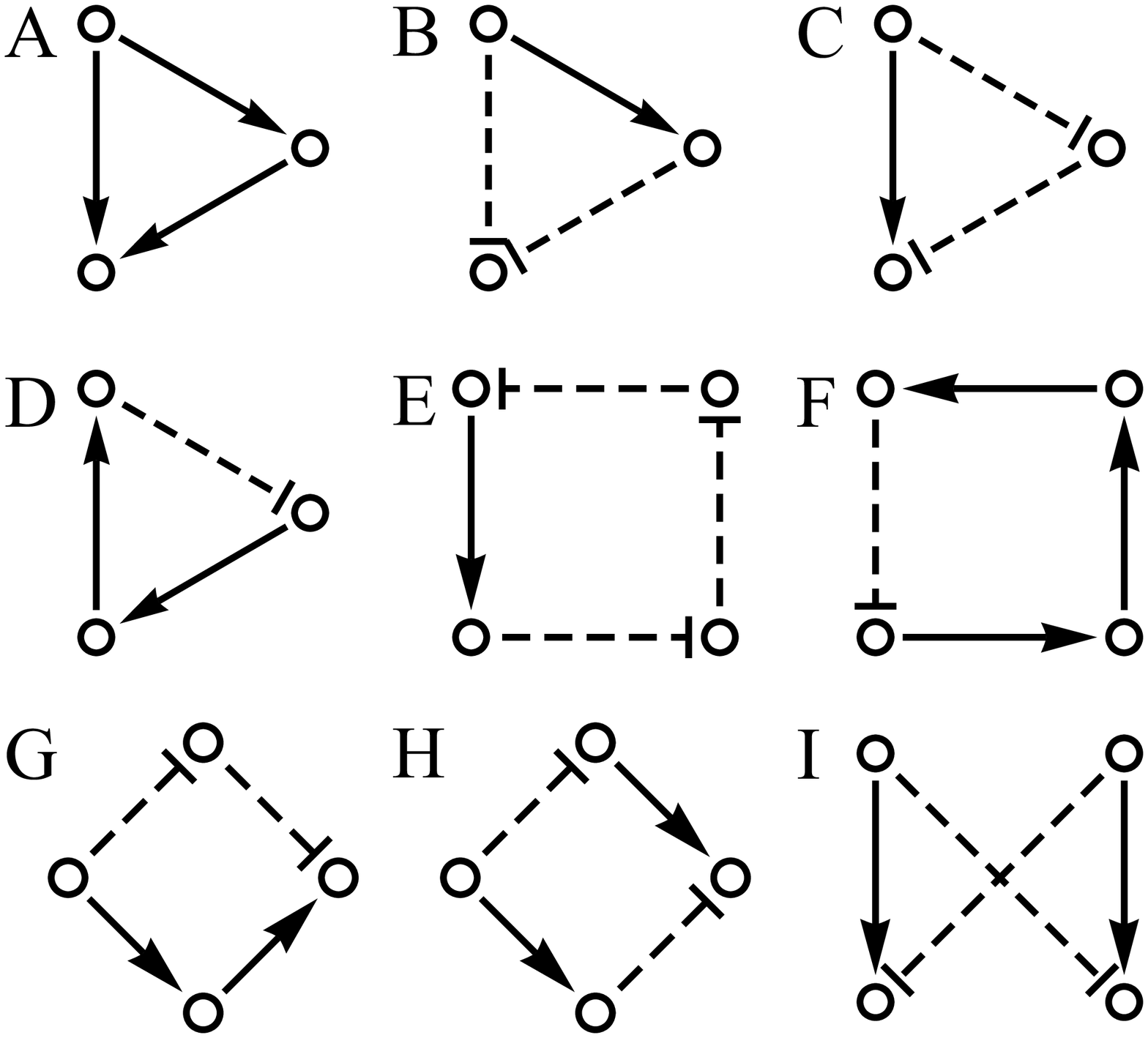}
\caption{Most prominent network motifs found from our GRN
ensembles. Corresponding frequencies are given in Tab.~\ref{tab:motifs}.}
\label{fig:motifs}
\end{figure}

Given the genotypes in each of our ensembles,
produced by the MCMC, we have extracted the motifs 
present in the associated essential networks.
The different motifs we find are represented graphically
in Fig.~\ref{fig:motifs}.
For two nodes, we find no significantly over 
expressed motifs.
For three nodes, we find that for both yeast cell cycle ensembles
several of the coherent feed-forward loops are 
over represented. In particular for the
baker's yeast ensemble, these over-representation factors are close to 10.
Finally, for four nodes, we find the presence of numerous 
motifs: two kinds of incoherent diamonds, two kinds 
of frustrated four-point loops and
incoherent bifans. The detailed frequencies of each
of the different motifs in these ensembles are
given in Tab.~\ref{tab:motifs}.

Finally, instead of working within the \emph{in silico}
GRNs produced by our ensembles, one can 
check for motifs in the wild-type networks.
We find that the motifs are not all the same as in the
associated ensemble. Among the strongest differences,
the RR motif consisting of two mutually inhibitory nodes
is strongly over-represented in the wild-type networks
but not in the \emph{in silico} networks. Differences
exist also for a number of other motifs. For example, 
in the wild-type network of baker's yeast, the numbers are:
motif RR = 3; A (C1-FFL) = 2, 
C (C4-FFL) = 1,  D (neg 3-loop) = 2. In the case of fission
yeast, the numbers are motif RR = 4; diamonds: G = 2, H = 2; 
bi-fan: I = 1. These numbers can be compared to those
in the GRN ensemble using Tab.~\ref{tab:motifs}.
\begin{table}[!ht]
\centering
\begin{tabular}{lccc}\hline \hline
  Motifs  &  Baker's  yeast &  Fission yeast &  Idealized cycle \\
 \hline
 A (C1-FFL)    & {\bf 1.04(2)  }&  0.0075(11) 	&  0.026(1)  \\
 randomized    & 0.30(1)        &  0.162(7)   	&  0.149(4)  \\
 B (C3-FFL)    & {\bf 1.93(2)  }& {\bf 0.23(1) }&  0.050(2)  \\
			   & 0.22(1)        & 0.094(5)      &  0.073(2)  \\
 C (C4-FFL)    & {\bf 1.93(2)  }& {\bf 0.21(2) }& {\bf 0.279(4)}\\
			   & 0.22(1)        & 0.099(5)      & 0.072(2)  \\
 D (neg 3-loop)& {\bf 0.80(1)  }&  0.091(6)   	&  0.0021(4) \\
			   & 0.34(1)        &  0.258(8)   	&  0.284(5)  \\
 E (neg 4-loop)& {\bf 0.81(2)  }& {\bf 0.063(5)}& {\bf 1.97(1)}\\
 		       & 0.061(3)       & 0.023(3)      & 0.029(2)  \\
 F (neg 4-loop)& {\bf 0.98(2)  }&  0.023(3)   	& {\bf 2.27(2)}\\
			   & 0.065(3)       &  0.020(2)   	& 0.029(2)  \\
 G (diamond)   & {\bf 0.41(2)  }& {\bf 0.098(5)}& {\bf 0.39(1)}\\
			   & 0.045(3)       & 0.030(4)      & 0.036(2)  \\
 H (diamond)   & {\bf 0.48(2)  }&  0.050(4)   	& {\bf 1.49(1)} \\
			   & 0.071(3)       &  0.050(3)   	& 0.111(3)  \\
 I  (bi-fan)   & {\bf 0.10(1)  }&  0.0037(9)  	& {\bf 0.73(1) } \\
			   & 0.023(3)       &  0.0035(8)  	& 0.0056(5) \\
  \hline \hline
 \end{tabular}
 \caption{Frequencies of motifs found in regulatory networks generated with 
baker's yeast, fission yeast and idealized cycle target patterns. For each 
motif, the first line provides the frequency in the MCMC ensemble and the 
second line provides the frequency in the randomized networks. Numbers in bold 
indicate motifs for which the frequency in the GRN ensemble is at 
least two times higher than in randomized ensemble. Motifs corresponding 
to the symbols are shown in Fig.~\ref{fig:motifs}.}
\label{tab:motifs}
\end{table}
%
\subsection{Regulatory logic}
\par
\begin{figure}
\includegraphics[width=6.0cm]{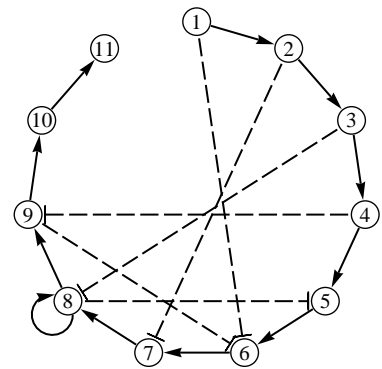}
\vspace{-0.8cm}
\caption{Most frequent essential network for the idealized cycle 
with 11 nodes. It arises in 6.1\% of all GRNs.}
\label{fig:MAM_GRN_freq}
\end{figure}

The pattern of edge usage obtained in Fig.~\ref{fig:freq_YST}
depends on the order of the genes. The order we chose
makes the successive expression states resemble as much as possible
a left to right shifting block of 1's. As is visible in 
Table~\ref{tab:pmb}, even with this ``best'' choice, irregularities
remain, suggesting one consider an idealized case. 
We thus replaced the irregular shifting block by
a regular one and then repeated our methodology on 
these idealized cell cycles. Interestingly, the 
associated essential networks have edge usage and motifs 
similar to the ones previously described for the wild-type 
cell cycles. Furthermore, 
the activating interactions typically form one long 
feed forward cascade. To illustrate this phenomenon,
we show in Figure~\ref{fig:MAM_GRN_freq} the most 
frequently found essential network
when idealizing the target sequence of Table~\ref{tab:pmb}. 
All activating (non-self) interactions
follow each other from ``START'' to the last expression state.
The function of such a cascade is clear but if one extracts an associated motif,
say three consecutive activating interactions, it has 
no autonomous dynamics. To understand
the functioning of such a motif, one has to 
externally drive it, providing inputs and initial
conditions that will initiate a ``block'' of on genes. 
The motif will then propagate as a cascade of 
falling dominoes~\cite{Murray_Kirschner_1989}
until the block of 1's is pushed out and the whole region consists
of 0's.
\par
\section{Discussion and conclusion}

It is quite remarkable that a model, based on
rather elementary thermodynamic and probabilistic
principles, succeeds in reproducing much of
the cell cycle network topology found in both
baker's yeast and fission yeast, two
highly divergent organisms. Our \emph{in silico} approach
is based on considering all possible networks subject to the
constraint that they give rise to the same gene expression
patterns as the experimentally observed systems. In effect,
this approach allows one to infer how network 
regulatory architectures are constrained by the network ``function''.
Within this framework, our modeling was able to give 
a good indication of the number of interactions
and of the actual edges that are used in the two
yeast species. We also considered higher order features such as
network motifs; this allowed us to reveal an ubiquitous design
principle in the form of an activatory cascade that is 
realized perfectly within an idealized 
cell cycle and is manifest in the less regular cell cycles studied here.

However in the case of the mammalian cell cycle network
our results are less impressive to say the least.
As already mentioned, this is not really a surprise.
In view of the simplicity of the model, its failures
are not devoid of interest: they indicate where a 
more sophisticated modeling is necessary. It is 
reassuring that the model does better for uni-cellular
organisms than multi-cellular organisms which have
far more levels of regulatory control; the contrary would be
suspect!
\par
We were seeking to determine which aspects of GRN
structure are expected from simple modeling and arguments of
a generic nature. Hence, we have assumed that all
interactions are a priori possible and only
allowed the
expression phenotype to constrain the networks. But it is
clear that ignoring chemistry in a biological
process is somewhat limiting and may be responsible
for some of the features arising in the biological networks
but not in our \emph{in silico} networks. A likely candidate
for this is the motif consisting of 2 mutually repressing 
molecular species: it is very frequent in the two
yeast cell cycle networks but not in our ensembles
of GRN. This absence occurs in spite of the fact that
for other phenotypic constraints, namely having
fixed point expression patterns, our model does produce
such motifs. Thus it would be of interest
to see the impact on our results of forbidding
interactions known to be unlikely on biochemical
grounds, but such a study is beyond the scope
of the present work. 

\par
{\bf Acknowledgments}
This work was supported by the International Ph.D.
Projects Programme of the Foundation for Polish Science
within the European Regional Development Fund of the
European Union, agreement no. MPD/2009/6. The LPT is
an Unit\'e de Recherche de l'Universit\'e Paris-Sud associ\'ee au CNRS.


\end{document}